\documentclass[conference,10pt]{IEEEtran}
\usepackage{epsfig,rotating,setspace,latexsym,amsmath,epsf,amssymb,amsfonts,bm,theorem,epstopdf,cite,mathtools,caption,subcaption,enumerate,longtable,accents,bbm,algorithm,graphicx,epsf,authblk,url,color,multirow,comment,tabularx,dsfont,physics,tikz}
\usepackage{mathtools, cuted}
\usepackage[inline]{enumitem}
\usepackage[noend]{algpseudocode}

\algrenewcommand\algorithmicforall{\textbf{foreach}}
\algrenewcommand\algorithmicindent{.8em}

\newcommand{\figref}[1]{\figurename~\ref{#1}}
\IEEEoverridecommandlockouts
\allowdisplaybreaks

\begin{document}

\title{Double Spending Analysis of Nakamoto Consensus for Time-Varying Mining Rates with Ruin Theory}

\author[1]{Mustafa Doger}
\author[1]{Sennur Ulukus}
\author[2]{Nail Akar}
\affil[1]{\normalsize Department of Electrical and Computer Engineering, University of Maryland, College Park, MD, USA}
\affil[2]{\normalsize Electrical and Electronics Engineering Department, Bilkent University, Ankara, T\"{u}rkiye}

\maketitle
\let\thefootnote\relax\footnotetext{This work is done when N.~Akar is on sabbatical leave as a visiting professor at University of Maryland, MD, USA, which is supported in part by the Scientific and Technological Research Council of T\"{u}rkiye  (T\"{u}bitak) 2219-International Postdoctoral Research Fellowship Program.}

\begin{abstract}
    Theoretical guarantees for double spending probabilities for the Nakamoto consensus under the $k$-deep confirmation rule have been extensively studied for zero/bounded network delays and fixed mining rates. In this paper, we introduce a ruin-theoretical model of double spending for Nakamoto consensus under the $k$-deep confirmation rule when the honest mining rate is allowed to be an arbitrary function of time including the block delivery periods, i.e., time periods during which mined blocks are being delivered to all other participants of the network. Time-varying mining rates are considered to capture the intrinsic characteristics of the peer to peer network delays as well as dynamic participation of miners such as the gap game and switching between different cryptocurrencies. Ruin theory is leveraged to obtain the double spend probabilities and numerical examples are presented to validate the effectiveness of the proposed analytical method. 
\end{abstract}

\section{Introduction}
Nakamoto consensus which is described in detail in \cite{btc-whitepaper} is the core technology of Proof-of-Work (PoW) blockchains. In PoW blockchains, the presence of adversaries and variable network delays among the participants of the blockchain may jeopardize the security of a transaction in the form of double spending, whose analysis is the focus of the current paper. We refer the reader to \cite{information_propagation, jiang_wu_BC21} for network delays in blockchains and their adverse impact on performance.  For more details on Nakamoto consensus and PoW blockchains, we refer the reader to \cite{bashir2020mastering, xiao2020survey, narayanan2016bitcoin}. The double spending problem arises as the outcome of a private double spending attack described below based on \cite{rosenfeld2014analysis}.  A transaction $tx$ is published through the network, where the adversary pays the merchant for a service or goods, and $tx$ is included in the next public block to be mined. The attacker then privately starts to mine a branch which includes a conflicting transaction $tx'$ which pays the attacker itself. While doing so, the attacker waits until the  transaction $tx$ is confirmed and the merchant sends the product before publishing its own branch. At this point, the attacker continues to mine on its private chain until it turns out to be at the same height as the honest chain, upon which the attacker broadcasts its private chain. As the outcome of the longest chain protocol, some participants will switch to the adversary chain, and consequently, the transaction $tx$ is compromised.
 
In the Nakamoto consensus, network delay refers to the amount of time needed until a mined block is successfully received by the other participants through the peer to peer network. Early work on double spending focused on the zero network delay case, fixed mining rates for both the honest and adversary chains, and the $k$-deep confirmation rule in which a block along with all the transactions inside it, is confirmed in the longest chain if at least $k-1$ blocks have been mined on top of it. A simple approximate analysis appeared first in \cite{btc-whitepaper} for the double spend probability by making use of the results for the gambler's ruin problem of applied probability \cite{feller1968introduction}. A more accurate analysis is provided in \cite{rosenfeld2014analysis}, and \cite{DoubleSpendRaces} presents a closed-form expression for the probability of success of a double spend attack using the regularized incomplete Beta function and asymptotic expressions as $k$ grows. However, note that as $k$ grows, transaction latencies also increase, known as the security-latency trade-off. The authors of \cite{PINZON201679} note that ignoring the time needed to mine $k$ blocks leads to inaccurate assessment of the double spending probabilities which are obtained in \cite{grunspan2020mathematics} by using sophisticated confirmation rules considering both $k$ and $t$.

Recently, there has been a surge of interest in the investigation of the security-latency trade-off in the presence of network delays\cite{nakamoto-always-wins, garay, gazi-consist-bounds, kiffer-method-analyze-consistency, pass-Analysis-blockchain}. In the so-called bounded delay model, all honest participants are guaranteed to receive a mined block after a bounded delay $\Delta$, for which the security-latency problem has been studied extensively in \cite{ guo-close-sec-lat, our-sec-lat-extended, cao2023tradeoff, gazi-sec-lat}. These models are concerned with giving theoretical guarantees, therefore, they consider the case where the network delay experienced by most honest blocks is exactly $\Delta$, which we call fixed-delay model in this paper. In reality, however, some participants will receive a mined block soon after the mining instance and they can start to mine on a legitimate block right away, whereas a fraction of the participants will receive the mined block relatively late, creating forks. Out of an experimental study of information propagation in the Bitcoin network, \cite{information_propagation} obtains an empirical probability density function of time until a node receives a block with a median of 6.5 seconds, mean of 12.6 seconds, and $5 \%$ of the nodes receiving the block after 40 seconds. Both \cite{information_propagation} and the raw data of the inventory messages publicly available at \cite{DSN-Bitcoin-Monitoring} suggest that simply assuming zero-delay or fixed-delay may not reflect the reality. 

There has also been a recent interest to analyze the Nakamoto consensus under random network delays \cite{pow-under-random-safe,our-queue-sec-ext-version,our-random-delay,parallel-pow-bounds}. The delay period is assumed to be exponentially distributed in \cite{our-queue-sec-ext-version} and the results are then transferred to queuing theory and connected to the throughput. \cite{our-random-delay} generalizes the pessimistic models above to any random delay distribution and connects the results to the mining gap \cite{gap-game, instability-no-block-reward}.

Another issue with fixed-delay results in \cite{nakamoto-always-wins, guo-close-sec-lat, guo-btc-sec-lat, our-sec-lat-extended, cao2023tradeoff, gazi-sec-lat} is that they ignore the effects of the adversarial strategies on the block interval adjustment rule. It is shown in \cite[Appendix~A]{Sompolinsky2015SecureHT} that the adversary should not publish any block before the attack. For example, if the adversary mines on a different cryptosystem temporarily, the puzzle difficulty decreases similar to the coin-hopping attacks \cite{coin-hopping-meshkov,profit_lag}. Then, when the adversary mounts the attack, during a delay interval, the number of adversarial blocks mined on new heights increases while honest miners are wasting their hashpower towards forks. As a result, forkrate and double spending probability increase.

Motivated by the work in \cite{information_propagation}, we attempt to model the honest mining rate during block delivery periods by an arbitrary function of time, which can be obtained from experimental data, to more accurately obtain the double spend probabilities. To the best of our knowledge, there is no work to model the honest mining rate as a function of time, as we do in this paper, for more accurate estimation of the double spend probabilities under variable network delays. However, the attacker's mining rate is assumed to be fixed in this paper and the balanced attacks that can effectively increase it during block delivery periods are left for future research. 

The contributions of this paper are as follows:
\begin{itemize}
\item Since the gambler's ruin framework in \cite{btc-whitepaper, rosenfeld2014analysis} is not directly applicable to our more general system model involving time-varying mining rates, we resort to the use of the more general ruin theory studied in detail in \cite{asmussen-ruin}, and more specifically the discrete ruin model of \cite{Dickson_2016}, to obtain the double spend probabilities. 
\item The ruin-theoretical model we propose in this paper requires the first $k$ probability masses as well as the mean, of the distribution of blocks mined by the adversary during an honest inter-mining period, which is denoted by the non-negative integer-valued random variable $\Phi$. We propose to fit a matrix exponential distribution to the honest inter-mining times when the honest mining rate is time-varying, which subsequently allows us to obtain the probability mass function (pmf) and the mean of $\Phi$, employing a matrix-analytical algorithm, that is computationally stable and efficient.
The impact of block interval adjustment rule has been ignored in \cite{guo-close-sec-lat, guo-btc-sec-lat, our-sec-lat-extended, cao2023tradeoff, our-queue-sec-ext-version, our-random-delay}. When the adversary does not publish any blocks temporarily, the block puzzles become easier and the number of blocks that can be mined during a delay interval increases which has an impact on double spending. Further, dynamic participation of honest miners also effects the block intervals and the double spending probabilities. We consider this effect for the first time in our model.
\end{itemize}

\section{Preliminaries} \label{sec:preliminaries}
\subsection{ME-fication and Approximating Deterministic Variables} \label{sec:cme}
Phase-type (PH-type) distributions are very commonly used for modeling in queuing systems \cite{neuts81, neuts2013,asmussen_etal_SJS96}. A Markov process is defined on the state-space $\mathcal{S} = \{1,2,\ldots,K,K+1\}$ with one absorbing state $K+1$, initial probability vector $v$ of size $1 \times K$, and an infinitesimal generator of the form, 
\begin{align} 
    \begin{pmatrix} 
    T & h \\
    \mathbf{0} & 0 \end{pmatrix},
\end{align}
where the sub-generator $T$ is $K \times K$, and $h$ is a column vector such that $h=-Te$, where $e$ is a column vector of ones. The diagonal elements of are strictly negative, and its off-diagonal elements are non-negative. The distribution of time till absorption into state $K+1$, is called PH-type characterized with the pair $(v,T)$, i.e., $X \sim PH(v,T)$ with order $K$. The cumulative distribution function (cdf) and the probability density function (pdf) of $X \sim PH(v,T)$ can be written as, 
\begin{align}
    F_X(x) & =1 -v \mathrm{e}^{Tx} e, \ 
    f_X(x)= -v \mathrm{e}^{Tx} T e, \ x\geq 0.
    \label{phdensity}
\end{align}
A generalization of the PH-type distribution is the so-called matrix exponential (ME) distribution \cite{AsmussenBladt97, fackrell_thesis, he_aap07} for which  $X \sim ME(v,T)$ with order $K$ when the pdf of the random variable $X$ is in the same form \eqref{phdensity} as in PH-type distributions, however, $v$ and $T$ do not possess the same stochastic interpretation. The moment generating function (mgf) $M_X(s)=\mathbb{E} [e^{sX}]$ of $X \sim PH(v,T)$ or $X \sim ME(v,T)$ are both written as,
\begin{align}
    M_X(s) & = -v (sI+T)^{-1} h, \label{general_form}
\end{align}
which is a strictly proper rational function of $s$ with degree $K$ and $h = - Te$. ME distributions can also be used in place of PH-type using the same computational techniques \cite{Horvath2016,Bean2010,BUCH09a,BUCH10b}. 

A deterministic variable, $X$, can be approximated by a ME distribution. In the Erlangization method \cite{asmussen.erlangian.2002, ramaswami.2008}, $X=\delta$ is approximated by an Erlang-$K$ distribution with order $K$, i.e., $X^{(K)} \sim PH(v^{(K)},T^{(K)})$, where $v^{(K)}  =  \begin{pmatrix} 1 & 0 & 0 & \cdots & 0 \end{pmatrix}$ and
\begin{align}
    T^{(K)}  = & \frac{K}{\delta} \begin{pmatrix}
    -1 & 1 & & & \\
     & \ddots & \ddots & \\
     & & -1 & 1 \\
     & & & -1
    \end{pmatrix}.
\end{align}
As the order $K$ increases, $X^{(K)}$ converges to $X=\delta$ in distribution but with relatively slow convergence rate since the squared coefficient of variation (scv) of $X^{(K)}$ is $\frac{1}{K}$.

\begin{figure}[t]
	\centering
	\includegraphics[width=\columnwidth]{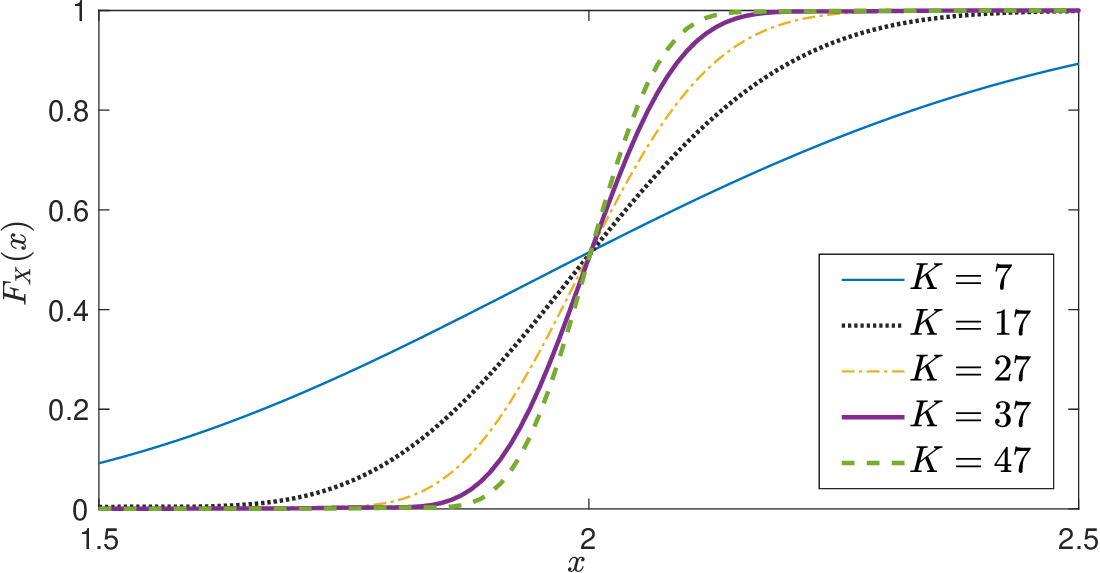}
	\caption{The cdf $F_{X}(x)$ obtained with the CME distribution with order $K$ for approximating deterministic variable $X=2$.}
	\label{fig:fig1}
\end{figure}

In the alternative ME-fication method, concentrated ME distributions (CME) are proposed in \cite{cme} for approximating deterministic quantities. The asymptotic behavior of the scv of the proposed CME distribution of order $K$ is $\frac{2}{K^2}$. Fig.~\ref{fig:fig1} depicts the cdf obtained with the approximate CME distribution, denoted by $\text{CME}[K,\delta]$ proposed in \cite{cme} for odd values of $K$ with rational mgfs when $\delta=2$. The accuracy of this approximation improves quadratically as $K$ increases. With the order of the approximating distribution being fixed, ME-fication provides better fits than Erlangization \cite{MEfication}.

\subsection{Probability Generating Function} \label{sec:pgf}
For convenience of notation, in addition to the usual definition of probability generating function (pgf) of $Y$, denoted by $G_Y(z)$ \cite{yates2014probability, alfa2015applied}, we use the $k$-partial pgf, denoted by $G^{(k)}_Y(z)=\{ G_Y(z) \}_k$, and defined as,
\begin{align}
    G^{(k)}_Y(z) & = \sum_{n=0}^{k-1} p_Y(n) z^n. 
\end{align}
When the pmf $p_Y(n)$ is in matrix geometric form,
\begin{align}
    p_Y(n) & = c A^n b, \quad n=0,1,\ldots \label{MatrixGeometric}
\end{align}
where $c$ is a row vector, $A$ is a real square matrix with all its eigenvalues lying inside the unit circle, and $b$ is a column vector, the pgf of $Y$ takes the following form,
\begin{align}
    G_Y(z) & = c (I + Az + A^2 z^2 + \cdots) b = c (I - Az)^{-1} b. \label{pgf2}
\end{align}
A Poisson random variable $Y$ with mean $\lambda$ has the pgf of the form $G_Y(z)=\mathrm{e}^{\lambda(z-1)}$. Assume events occur according to a Poisson process with parameter $\lambda$ in a random amount of time $X$. Given $X=x$, $Y$ is Poisson distributed with mean $\lambda x$ and 
\begin{align}
    G_Y(z) & = \int_{x=0}^{\infty} \mathrm{e}^{\lambda x(z-1)} f_X(x) \dd{x} = M_X(\lambda (z-1)). \label{NumEventsRandomTime}
\end{align}

\subsection{Lindley Equation} \label{sec:lindley}
We consider the discrete-time Lindley process $Q_i$, $i=0,1,\ldots,$ governed by the following Lindley equation \cite{Lindley_1952},
\begin{align}
    Q_{i+1} & = (Q_i +\Phi_i -1)^+ = \max (Q_i +\Phi_i -1,0), 
    \label{lindley}
\end{align}
where $\Phi_i$s, $i=1, 2,\ldots,$ are i.i.d.~non-negative random variables with common pmf $p_{\Phi}(n)$, cdf $F_{\Phi}(n)$, and complementary cdf (ccdf) $\bar{F}_{\Phi}(n)$. Assuming $\mathbb{E}[\Phi] < 1$, the pmf of the limiting random variable $Q$ exists (see \cite{alfa2015applied}) and is given by the following iterations (see \cite{our-random-delay} for a derivation),
\begin{align}
    p_Q(0) & = \lim_{i \rightarrow \infty} \mathbb{P}(Q_i = 0)= \frac{1-\mathbb{E} [\Phi]}{p_{\Phi}(0)}, \\ 
    p_Q(n) & =  \lim_{i \rightarrow \infty} \mathbb{P}(Q_i = n)= \frac{\sum_{j=0}^{n-1}p_Q(j) \bar{F}_{\Phi}(n-j)}{p_{\Phi}(0)}, \ n \geq 1.
    \label{lindley22}
\end{align}

\subsection{Discrete-Time Risk Problem} \label{sec:ruin}
 Recently, insurance risk problems were used in finding probabilities of successful withholding attacks and associated profits in the blockchain context \cite{goffard_insurance_selfish,Goffard_Fraud_risk}. Here, we consider a specific discrete-time insurance risk problem studied in \cite{Dickson_2016}, i.e., the discrete-time surplus process $U_n$, $n=1, 2,\ldots$,
\begin{align}
    U_{n} & = u + n - \sum_{i=1}^n \Phi_i, \label{dickson}
\end{align}
where $u=U_0$ is the insurer’s initial surplus, $\Phi_i$ denotes the insurer’s aggregate claim amount in interval $i$, $\Phi_i \sim \Phi$ are i.i.d.~taking non-negative integer values. The insurer’s premium income in one interval is one. Ultimate ruin occurs if the surplus ever falls to zero or below. Let the random variable,
\begin{align}
    \tau(u) & = \inf \{ n \geq 1: \ U_n \leq 0 \},  
\end{align}
be the first epoch when the surplus falls to zero or below. The so-called ultimate ruin probability, denoted by $\psi(u)$, is then,
\begin{align}
    \psi(u) &= \mathbb{P}( \tau(u) < \infty), \\ 
            &= \mathbb{P} \bigg( u + n - \sum_{i=1}^n \Phi_i \leq 0, \text{ for some } n \geq 1 \bigg). 
\end{align}
The following simple expressions give $\psi(u)$,
\begin{align}
    \psi(0) & = \mathbb{E} [\Phi], \\
    \psi(u) & = \mathbb{E} [\Phi] + \sum_{j=0}^{u-1} \bar{F}_{\Phi}(j) \left( 
    \psi(u-j)-1 \right), \quad u \geq 1. \label{ruin1}
\end{align}
Following \cite{our-random-delay}, for $i\geq 1$, the ruin probabilities are also related to the steady-state solution of the Lindley equation in \eqref{lindley22},
\begin{align}
    \psi(u) & =\bar{F}_{Q}(u-1)=1-\mathbb{P}(Q<u), \quad u \geq 1. \label{ruin2}
\end{align}

\section{System Model} \label{sec:system}
Let $\alpha$ (resp.~$\beta$) denote the honest miners' full hashrate which follows the longest chain protocol (resp.~hashrate of the adversary). We consider a single adversarial entity with a fixed hashrate representing a collection of fully-coordinated adversaries, which does not have to adhere to the longest chain protocol and ties are broken in its favor, and becomes aware of honest blocks immediately when they are mined. When the honest block $i$, $i \geq 1$, is mined, it takes $\Delta$ time for the mined block to be disseminated to all the honest participants of the network. During this time, miners who do not receive the new block, can create forks, which do not increase the length of the longest chain, hence, we ignore them. However, miners who receive the block before the maximum delay $\Delta$ is reached, may start mining on the received block upon reception. In order to model this general scenario, we define a hashrate function $\alpha(t)$, $t \geq 0,$ that stands for the effective hashrate of the honest chain at time $t$ following a newly mined block at time zero. Note that $\alpha(t)=\alpha$ for $t \geq \Delta$ and the hashrate function $\alpha(t)$ is a general monotonically increasing function of $t$ in the closed interval $t \in [0,\Delta]$ which can be obtained from experimental data (see Section~\ref{sec:numerical}).

We consider the pre-mining attacks \cite{sompolinsky2016bitcoin} where the arrival of $tx$ is random and the adversary mines a private block whenever the private branch is longer than the honest branch. When the honest chain is longer, then the attacker starts a new branch on the honest chain until the arrival of $tx$.  Therefore, there is a possibility that the adversary chain may be longer than the honest chain when the transaction $tx$ appears in the mempool, which we call the lead in this paper, and find its distribution. 

Let $\Theta_i$ denote the inter-mining times of honest blocks $i$ and $i+1$; see Fig.~\ref{fig:SystemModel}. We assume that $\Theta_i$s are i.i.d.~and $\Theta_i \sim \Theta$ with pdf $f_{\Theta}(x)$, $x \geq 0$, and corresponding mgf $M_{\Theta}(s) = \mathbb{E}[\mathrm{e}^{s \Theta}]$. Also, let $\Phi_i$ denote the number of blocks mined by the adversary during the inter-mining time $\Theta_i$. Since the mining rate of the adversary is fixed in this paper, $\Phi_i \sim \Phi$ with pmf $p_{\Phi}(n)$, $n=0,1,\ldots$ and corresponding pgf $G_{\Phi}(z) = \mathbb{E}[z^{\Phi}]$. Clearly, the network delay $\Delta$ and the hashrate function $\alpha(t)$, $0 \leq t \leq \Delta$, have impact on the distribution of $\Phi$.

We assume $\mathbb{E}[\Phi]<1$, a $\mathcal{T}$-block interval rule and the $k$-deep confirmation rule. The condition $\mathbb{E}[\Phi]<1$ makes sure that the double spending probability is not trivially $1$. $\mathcal{T}$-block interval rule adjusts the puzzle difficulty such that the average time for mining a block remains $\mathcal{T}$ seconds on average. According to the $k$-deep confirmation rule, a block (along with all the transactions inside it) is said to be confirmed if it is part of the longest chain and at least $k-1$ blocks have been mined on top of it. The safety violation, or double spending, refers to the event that a transaction $tx$ is confirmed according to the $k$-deep confirmation rule in the view of an honest miner and later gets replaced (or discarded) with another conflicting transaction $tx'$ in the same honest view or another honest view.

\section{The Analytical Method} \label{sec:analytical}
To obtain the double spend probabilities, the first step is to find the mgf $M_{\Theta}(s)$ of the honest inter-mining times. 

In the zero-delay model, $\Delta=0$, once block $i$ is mined, the full hashrate $\alpha$ is used to mine block $i+1$. Hence, block inter-arrival times are exponentially distributed with parameter $\alpha$, giving rise to the following rational mgf,
\begin{align}
    M_{\Theta}(s) & = \frac{\alpha}{\alpha -s},
\end{align}
which can be written in the form \eqref{general_form} with $T=-\alpha$ and $v=1$.

In the fixed-delay model, the propagation delay is fixed as $\Delta$, and all miners receive the block after $\Delta$ time. During the delay, the honest mining power is assumed to be non-usable for mining a new block, giving rise to the following mgf $M_{\Theta}(s)$,
\begin{align}
    M_{\Theta}(s) & = \mathrm{e}^{\Delta s} \ \frac{\alpha}{\alpha -s}
\end{align}
which is not rational. Thus, we propose to use the $\text{CME}[K,\Delta]$ ME-fication method which uses a CME $X \sim ME(v^{(1)},T^{(1)})$ so that $M_{\Theta}(s)$ is written in the general form \eqref{general_form} with,
\begin{align}
    v & =  \begin{bmatrix} v^{(1)} & 0 \end{bmatrix},  \label{yeni1} \\
    T & = \left[ \begin{array}{c|c}
        T^{(1)} & h^{(1)} \\ \hline
        0 & -\alpha
    \end{array} \right], \label{yeni2} \\
    h^{(1)} & = -T^{(1)} e. \label{yeni3}
\end{align}

For the exponential delay model of \cite{our-queue-sec-ext-version}, the network delay is assumed to be between $[0,\Delta_h]$ for blocks at height $h$ with $\Delta_h$, $h \geq 1$ being i.i.d.~exponential random variables. For the case of $\Delta_h$ having a general ME distribution characterized with the pair $\Delta_h \sim ME(v^{(1)},T^{(1)})$, then $M_{\Theta}(s)$ is written in the form \eqref{general_form} where the parameters of this expression, namely $v,T$ and $h$ are given as in \eqref{yeni1}, \eqref{yeni2} and \eqref{yeni3}, respectively.

In the variable-delay model, a general hashrate function $\alpha(t)$, $0 \leq t \leq \Delta$, is employed to approximate the effective hashrate of the honest chain at time $t$ following a mined block at time zero. Note that $\alpha(t)=\alpha$ for $t \geq \Delta$, since all participants come to know about the mined block after the duration $\Delta$. For tractability purposes, we focus our attention to a piece-wise constant hashrate function with $N+1$ thresholds $0 = \Delta_0 < \Delta_1 < \Delta_2 < \cdots < \Delta_N=\Delta$, such that $\alpha(t)=\alpha_i$ when $\Delta_{i-1} \leq t < \Delta_{i}$, $1 \leq i \leq N$. We note that, it is always possible to approximate arbitrary time functions with piece-wise constant functions to any desired accuracy. Subsequently, for each interval $[\Delta_{i-1},\Delta_i)$ with length $\delta_i = \Delta_{i}-\Delta_{i-1}$, $1 \leq i \leq N$, we use the method $\text{CME}[K,\delta_i]$ described in Subsection~\ref{sec:cme} to construct $X^{(i)} \sim ME(v^{(i)},T^{(i)})$ with $h^{(i)} = -T^{(i)} e$. As a result, the random variable $\hat{\Theta} \sim ME(v,T)$ can be used to approximate the distribution of the actual inter-mining time $\Theta$ where $v$, $T$, and $M_{\Theta}(s)$ are given in \eqref{Big1}, \eqref{Big2} and \eqref{Big3}, respectively,
where the order of the ME distribution is $m=N K+1$, and $\tilde{T}^{(i)}$ denotes $T^{(i)}-\alpha_i I$, 
\begin{align}
    &v  =  \begin{bmatrix} v^{(1)} & \bm{0}  & \bm{0} & \cdots & \bm{0} & 0 \end{bmatrix}, \label{Big1}  \\
    &T  = \!\!\left[ \begin{array}{l|l|l|l|l|l}
        \!\tilde{T}^{(1)}\!\!\!& \!h^{(1)} v^{(2)}\!\!\!& & & & \\ \hline
        &  \!\tilde{T}^{(2)}\!\!& \!h^{(2)} v^{(3)}\!\!\!& & & \\ \hline
        & & \ddots & \ddots & & \\ \hline
        & & & \!\tilde{T}^{(N-1)}\!\!& \!h^{(N-1)} v^{(N)}\!\!\!& \\ \hline
        & & & & \!\tilde{T}^{(N)}\!\! & \!h^{(N)} \!\!\!\\ \hline
        & & & & & \!-\alpha\!\!\! \\
    \end{array} \right]\!\!, \label{Big2} \\
    &M_{\Theta}(s) = - v (sI+T)^{-1} h, \quad h=-Te. \label{Big3}
\end{align}

If the adversary temporarily works on another system before it mounts the attack, $\mathbb{E}[\Theta]=\mathcal{T}$ needs to be satisfied where only $\alpha(t)$ determines the puzzle difficulties. Note that,
\begin{align}
    \mathbb{E}[\Theta] = \dv{M_X(s)}{s} \bigg|_{s=0}=-vT^{-1}e.
\end{align}
A simple method to achieve $-vT^{-1}e=\mathcal{T}$ is to first assume $\alpha=1/\mathcal{T}$, find associated $T$ and $\mathbb{E}_\Theta$, then repeat with $\alpha\coloneqq\alpha\times\mathbb{E}_\Theta/\mathcal{T}$ accordingly until a predetermined accuracy (we pick $10^{-4}$) is reached. It is also possible to use bisection search on $\alpha$. We assume that $M_{\Theta}(s)$ is given as in \eqref{Big3}, which covers all the scenarios described in this section, namely zero-delay, fixed-delay, random-delay and variable-delay models.

Next, the pgf of $\Phi$, $G_{\Phi}(z)$, can be written from \eqref{NumEventsRandomTime},
\begin{align}
   G_{\Phi}(z) &= M_{\Theta}(s) =  \left( -v (sI+T)^{-1} h \right) \bigg|_{s=\beta(z-1)} \\
    & = c(I-Az)^{-1} b, 
\end{align} 
where 
\begin{align}
  A & = \left(I - \frac{T}{\beta}\right)^{-1}, \quad c= \frac{vA}{\beta}, \quad b=h,
\end{align}
with the square matrix $A$ being of size $m$. From \eqref{pgf2}, the pmf $p_{\Phi}(n)$ can then be written as in \eqref{MatrixGeometric}. Thus, the first $k$ probability masses can be obtained by the following, 
\begin{align}
    y_n&=Ay_{n-1}, \quad n=1,2,\ldots,k-1, \ y_0=b, \\ 
    p_{\Phi}(n) & =c y_n, \quad  \quad \ n=0,1,\ldots,k-1. 
\end{align}
Consequently, the $k$-partial pgf of $\Phi$ can be written as,
\begin{align}
    G^{(k)}_{\Phi}(z) & = \sum_{n=0}^{k-1} p_{\Phi}(n) z^n. 
\end{align}
We will work with the complementary event of double spending, hence, we are not interested in the rest of the probability masses $p_{\Phi}(n),n\geq k$. The mean of $\Phi$ can be written as,
\begin{align}
    \mathbb{E} [\Phi]  & = c A (I-A)^{-2} b. 
\end{align}

\begin{figure*}[t]
    \centering
    \begin{tikzpicture}[scale=0.20]
    \draw[very thick,|->] (0,10) -- (80,10);
    \draw (80,10) node[anchor=north] {$t$} ;
    \draw[very thick,|->] (0,6) -- (14,6);
    \draw (7,6) node[anchor=north] {{\scriptsize Pre-mining phase}} ;
    \draw[very thick,<->] (14,6) -- (60,6);
    \draw (37,6) node[anchor=north] {{\scriptsize Confirmation interval}} ;
    \draw[very thick,<-] (60,6) -- (80,6);
    \draw (70,6) node[anchor=north] {{\scriptsize Post-confirmation race}} ;
    \draw[thick,|->] (0,12) -- (7,12);
    \draw (3.5,12) node[anchor=south] {$\Theta_0$} ;
    \draw (10.5,12) node[anchor=center] {$\ldots$} ;
    \draw[thick,dotted] (14,16) -- (14,9.5);
    \draw[thick,<->] (14,12) -- (20,12);
    \draw (17,12) node[anchor=south] {$\Theta_r$} ;
    \draw[thick,dotted] (20,16) -- (20,9.5);
    \draw[thick,<->] (20,12) -- (33,12);
    \draw (26.5,12) node[anchor=south] {$\Theta_{r+1}$} ;
    \draw[thick,<->] (33,12) -- (43,12);
    \draw (38,12) node[anchor=south] {$\Theta_{r+2}$} ;
    \draw (14,10) node[anchor=north] {$\tau_{0}$} ;
    \draw[thick,dotted] (33,16) -- (33,9.5);
    \draw[thick,dotted] (43,16) -- (43,9.5);
    \draw (20,10) node[anchor=north] {$\tau_1$} ;
    \draw (33,10) node[anchor=north] {$\tau_2$} ;
    \draw (43,10) node[anchor=north] {$\tau_3$} ;
    \draw (48,12) node[anchor=center] {$\ldots$} ;
    \draw[thick,<->] (52,12) -- (60,12);
    \draw (56,12) node[anchor=south] {$\Theta_{r+k-1}$} ;
    \draw[thick,dotted] (60,16) -- (60,9.5);
    \draw (60,10) node[anchor=north] {$\tau_k$} ;
    \draw[thick,dotted] (74,16) -- (74,9.5);
    \draw (74,10) node[anchor=north] {$\tau_{k+1}$} ;
    \draw[thick,<->] (60,12) -- (74,12);
    \draw (67,12) node[anchor=south] {$\Theta_{r+k}$} ;
    \draw (77,12) node[anchor=center] {$\ldots$} ;
    \end{tikzpicture}
    \caption{Illustration of the honest chain view.}
    \label{fig:SystemModel}
\end{figure*}
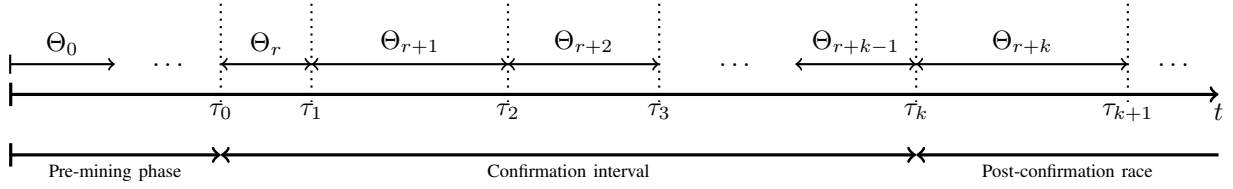

The double spending probability is described in line with the diagram given in Fig.~\ref{fig:SystemModel}. We assume that the particular transaction $tx$ is included in honest block at height $r+1$, $r\gg1$, which is mined at time $\tau_1$, where $\tau_{0}$ denotes the epoch where the honest block at height $r$ is mined. We also let $\tau_{i}+\Delta$, $i=1,2,\ldots$ denote the time when the block containing $tx$ is $i$-deep in all honest views. Once the transaction $tx$ appears at time $t$, $\tau_{0} < t < \tau_1$, the adversary starts to mine only on its own private chain. According to the $k$-deep confirmation rule, at time $\tau_{k}+\Delta$,  $tx$ is said to be confirmed in all honest views.

Let $H^v_{t}$ denote how deep the $tx$ is in the longest chain from the view of an honest miner $v$ at time $t\geq \tau_0$. Let $A_{t}$ denote the corresponding depth at the private chain at time $t$, i.e., how many blocks the private chain contains starting (excluding) from the height $r$. Let $Z^v_{t}=H_{t}^v-A_t$ denote the honest lead at time $t$ according to the honest miner $v$. Then, we let the random variable $\sigma$ denote the first such epoch 
\begin{align}
    \sigma & = \inf \{t: \inf_{v}Z^v_{t} \leq 0, \ t \geq \tau_k   \}.
\end{align}
The double spend (the safety violation) probability $p$ is then,
\begin{align}
p & = \mathbb{P}(\sigma < \infty). \label{DoubleSpend}
\end{align}

Let $t^-$ denote the time just before the instant $t$. Using \cite[Lemmas~1-2]{our-random-delay}, we have,
\begin{align}
    \inf_{t\geq\tau_k}\inf_{v}\{H^v_t-A_t\}&=\inf_{i\geq 1}\inf_{v}\{H^v_{(\tau_{k+i}+\Delta)^-}-A_{(\tau_{k+i}+\Delta)^-}\}\\
    &=\inf_{i\geq 1}\{k+i-1-A_{(\tau_{k+i}+\Delta)^-}\}\\
    &=\inf_{i\geq 1}\{Z_{i}^-\},
\end{align}
since $A_t$ is monotone increasing and $H_t^v$ has jumps at time instances ${(\tau_{k+i}+\Delta)}$ for some honest view $v^*$ which is yet to observe the honest block at height $r+k+i$. The honest view $v^*$ has the same function as the weakest possible view explained in \cite[Section~III.A]{cao2023tradeoff}. Thus, we can focus on safety violations at the embedded instants $(\tau_{k+i}+\Delta)^{-}$, $i\geq 1$, as opposed to arbitrary time points. For this purpose, we let the random variable $\kappa$ denote the first such epoch  
\begin{align}
    \kappa & = \inf \{j: \ Z_j^- \leq 0, \ j > 0  \}.
\end{align}
The safety violation probability $q$ at these epochs is then,
\begin{align}
    q & = \mathbb{P}(\kappa < \infty). \label{q}
\end{align}
A safety violation at an arbitrary time $t$ immediately implies a safety violation at the next embedded instant. Therefore, we conclude that the quantities $p$ and $q$ are the same. Thus, we will focus our attention to obtaining the probability $q$ in \eqref{q}. 

Similar to \cite{guo-close-sec-lat, guo-btc-sec-lat, our-sec-lat-extended, our-queue-sec-ext-version, cao2023tradeoff, our-random-delay}, we divide our analysis into three phases: pre-mining phase, confirmation interval and post-confirmation race. In the pre-mining phase, the adversary tries to build a private chain to gain some advantage (lead) over the honest chain. When the lead is less than or equal to zero, the adversary will always build a new private block on top of the honest chain in the pre-mining phase, otherwise, it will continue mining on its own private chain. Once the pre-mining phase is over, the adversary will always build on its own private chain. Here, we note a subtle technical difference in our analysis from the existing works. There, the authors assume to observe the lead at a particular instant $t$ when $tx$ arrives at the mempool, and then start the race of mining again. As block arrivals are modeled exponential, inspection paradox gives rise to an over-estimation of the double spending probability in those papers as they consider worst case scenarios and want to give theoretical upper bounds. As our goal here is to provide results with more realistic settings, we consider the renewal interval containing $t$ is the same as any other interval.

Let $Q$ denote the lead of the adversary at $\tau_{0}^+$ when the final block at height $r$ that does not contain transaction $tx$. Consider the following Lindley process,
\begin{align}
    Q_{i+1} & = (Q_i + \Phi_i -1)^+, \quad i=0,1,\ldots \label{lindley2}
\end{align}
where $\Phi_i$ denotes the number of blocks mined by the adversary in interval $\Theta_i$ with the same common pmf $p_{\Phi}(\cdot)$. The Lindley process \eqref{lindley2} models the lead of the adversary right after the honest mining instances, i.e., at the end of the interval $\Theta_i$, starting from the genesis block $1$. Since $r\gg 1$, the limiting distribution of the Lindley equation \eqref{lindley2} gives us the $k$-partial pgf of $Q$, denoted by $G^{(k)}_{Q}(z)$ using the method detailed in Subsection~\ref{sec:lindley} using the $k$-partial pgf and also the mean of $\Phi$. Now, let $V$ denote the lead of the adversary chain at time $(\tau_{k}+\Delta)^-$ with respect to the honest chain of height $r$. From Fig.~\ref{fig:SystemModel}, it is clear that the pgf of $V$ can be written as,
\begin{align}
    G_V(z) & = G^{(k)}_Q(z) G^{(k)}_{\Phi}(z)^{k} G^{(k)}_{\Delta}(z),
\end{align}
where $G_{\Delta}(z)$ denotes the pgf of the Poisson distributed random variable with mean $\Delta\beta$. Consequently,
\begin{align}
    G^{(k)}_V(z) & = \Big\{ G^{(k)}_Q(z) G^{(k)}_{\Phi}(z)^{k} G^{(k)}_{\Delta}(z) \Big\} _k = \sum_{n=0}^{k-1} p_V(n) z^n.
\end{align}
Let $Z=k-1-V$ denote the honest chain's lead (according to the minimum view $v^*$) with respect to the adversary chain both at time $(\tau_{k}+\Delta)^-$. Therefore,
\begin{align}
    p_Z(i) & =p_V(k-1-i), \quad i=0,1,\ldots,k-1.
\end{align}
Recall that $Z_j^-$ stands for the lead of the (minimum view) honest chain at embedded instances which is governed by the following random walk,
\begin{align}
    Z_n^- & = Z + n - \sum_{i=1}^n \Phi_{r+k+i},  \quad n \geq 1.
\end{align} 
Using the results of Subsection~\ref{sec:ruin}, the safety violation probability $q$ in \eqref{q} can be written as,
\begin{align}
    q & = \mathbb{P}(\kappa < \infty) \\ 
      & = \sum_{u=0}^{k-1} \mathbb{P}(\kappa < \infty| Z=u) \mathbb{P}(Z=u) + \mathbb{P}(Z < 0) \\
      & = 1 - \sum_{u=0}^{k-1} p_Z(u) (1 - \psi(u)),
\end{align}
and the ruin probability $\psi(u)$ is to be obtained by \eqref{ruin1} or \eqref{ruin2}.

\subsection{Reduction to Other Delay Models} \label{sec:reduction-section}
Our analysis can be reduced to lower bounds of the security-latency models under different network delay assumptions \cite{guo-btc-sec-lat, our-queue-sec-ext-version, our-random-delay} studied in the literature. We list them below.
\begin{enumerate}
    \item Setting $N=0$, $T=-\alpha$, $v=1$ results in \textit{zero-delay} model.
    \item Setting $N=1$ and $\alpha_{1}=0$ reduces the system to the \textit{fixed-delay} strategy of the security-latency analysis provided in \cite{our-sec-lat-extended}. The difference from our method is that, the parameters above give an ME-approximation of deterministic $\Delta$ of \cite{our-sec-lat-extended} where the accuracy of this approximation is improved as $K \rightarrow \infty$. 
    \item Picking $K=1$, $N=1$, $\alpha_{1}=0$, $T^{(1)}=-\mu_1$, and $v^{(1)}=1$ reduces our proposed system system to the $\Delta$ delay strategy of the security-latency analysis provided in \cite{our-queue-sec-ext-version} where network delay is assumed to be between $[0,\Delta_h]$ for blocks at height $h$ with $\Delta_h$ i.i.d.~exponentially distributed random variables with parameter $\mu_1$. We call this model \emph{exponentially-fixed-delay} in this paper.
    \item Although the analysis provided in \cite{our-random-delay} considers $\Delta$ delay strategy for any random $\Delta_h$ distribution, our analysis here can be reduced to those that involve a matrix exponential distribution to model $\Delta_h$. For example, our analysis can be used for the numerical analysis provided in \cite{our-random-delay}, where $\Delta_h$ is assumed to be following a right-skewed delay distribution (Erlang-$k$) with $\Delta_h\sim E_2(1)$.
\end{enumerate}

Note that none of the existing works consider scaling the mining rate in order to make sure $\mathcal{T}$-block interval rule is satisfied. Finding the honest block generation rate in rigged model of \cite{guo-btc-sec-lat} is too complex. On the other hand, a simple adjustment of $\lambda$ (resp. $\mu_m$) as $\frac{1}{(1-\beta)(\mathcal{T}-\Delta)}$  in \cite{our-sec-lat-extended} (resp. \cite{our-random-delay})  would make sure $\mathcal{T}$-block interval rule is satisfied. Similar adjustments can be made for \cite{our-queue-sec-ext-version}. However, these models also involve inspection paradox, hence, when we compare our results with these models, we simply take the corresponding ME-approximation with the reduction methods. Thus, in the numerical examples, whenever we refer to the zero-delay or fixed-delay model, we obtain their related results using the reduction of $T$-matrix instead of using results directly from \cite{guo-btc-sec-lat,our-sec-lat-extended} which do not originally obey $\mathcal{T}$-block interval rule.

\section{Numerical Results} \label{sec:numerical}
For the numerical results, we use the raw data provided in \cite{DSN-Bitcoin-Monitoring} for the mean block proporation times in the time period from 01/01/2019 to 12/17/2023 spanning almost 5 years, where each day contains average propagation delay of $6000$ to $10000$ nodes (data for some days are missing). We refer to the data spanning from 01/01/2019 to 12/17/2023 as \textit{full range data} and from 01/01/2023 to 12/17/2023 as \textit{congested network data} since the block propagation times are higher on average compared to the rest of the data. 

We perform the following processing on raw data in several steps. Some nodes in the data seem to send inventory messages much later than usual. We assume, they are light nodes that only echo messages and do not contribute to mining. Thus, in the first step, for a given parameter $\epsilon>0$, we find the $100(1-\epsilon)$-th percentile, or simply $\Delta(\epsilon)$ which is the cutoff delay that represents the largest of the block delays that occur $(1-\epsilon)$ of the time in the raw data, and subsequently eliminating all the block delays strictly larger. This step is needed to avoid overly skeptical estimates for the double spending probability. 

Let $M$ denote the size of data at the end the first step. In the second step, we partition the data into $N=N'+2$ bins; the first bin, denoted by $B_0$ contains reports of delay less than $1$ millisecond (usually this is the node that has mined the block, or multiple nodes belonging to the pool that mined the block), their proportion in the data is denoted by $\vartheta$. We assume that these nodes experience a delay of $1$ millisecond, which could be regarded as the time needed to prepare the next block without a nonce. 

The next bin denoted by $B_1$, contains the smallest $\lfloor M'/N' \rfloor$ block delays, $B_2$ contains the next $\lfloor M'/N' \rfloor$ block delays, and so on, and finally $B_{N}$ contains the remaining largest block delays, where $M' = M-B_0$. We then calculate the sample mean $b_i$ of all the block delays in bin $B_i$, $0 \leq i \leq N$ which are suggested to be equal to the thresholds $\Delta_i$s (described in Section~\ref{sec:analytical}) according to the following identity,
\begin{align}
    \Delta_i = b_{i-1}, \quad 1 \leq i \leq N.
\end{align} 
The parameter $N$ is then used in the discretization of the hashrate function $\alpha(t)$ for which purpose we propose to use the following expression for obtaining $\alpha_i = \alpha \bar{\alpha}_i$, $1 \leq i \leq N$,
\begin{align}
    \bar{\alpha}_1 & =0, \quad \bar{\alpha}_2 = \vartheta, \\
    \bar{\alpha}_i  & = \bar{\alpha}_{i-1}+ \frac{\lfloor M'/N' \rfloor}{M}, \quad 3 \leq i \leq N,  
\end{align}
whose use in the analytical method is detailed in Section~\ref{sec:analytical}. This method assumes that all miners have the same hashrate as we do not have the data showing the hash powers associated with inventory messages. Our model can be modified to study more realistic scenario of miners having different network delays and hashrates, which is left for future research. Next, we fit an ME distribution characterized with the pair $(v,T)$ of size $NK+1$, where $K$ is the order of the ME distribution used for each bin according to \eqref{Big1}, \eqref{Big2} and \eqref{Big3}. 

\begin{figure}[t]
     \centering
     \captionsetup[subfigure]{aboveskip=0pt,belowskip=4pt}
     \begin{subfigure}[b]{0.9\columnwidth}
         \centering
         \includegraphics[width=\textwidth]{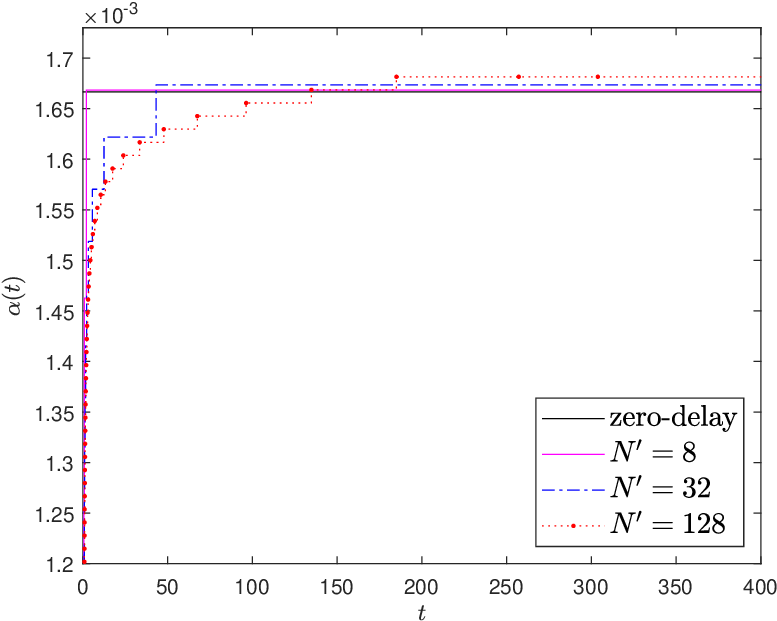}
         \caption{full range data}
         \label{fig:fullhash}
     \end{subfigure}
     \hfill
     \begin{subfigure}[b]{0.9\columnwidth}
         \centering
         \includegraphics[width=\textwidth]{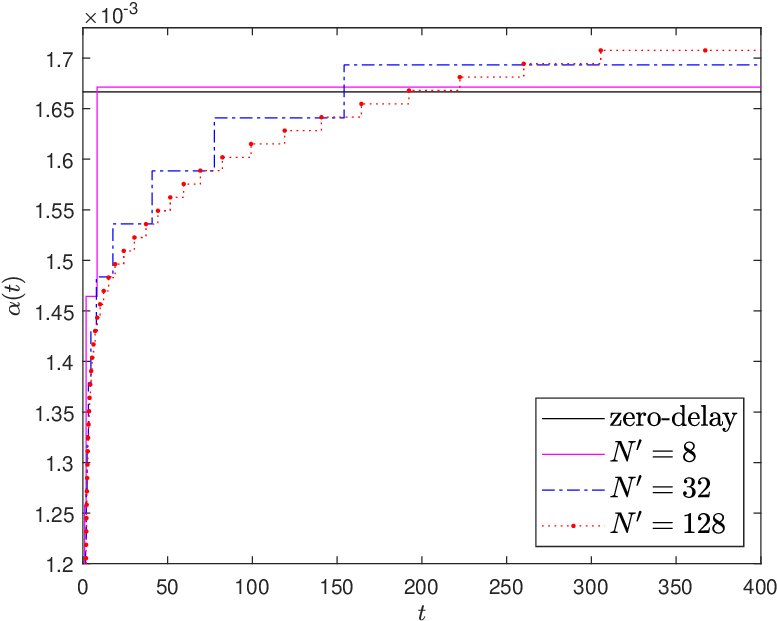}
         \caption{congested network data}
         \label{fig:conghash}
     \end{subfigure}
     \caption{The hashrates $\alpha(t)$ of the proposed ME distribution with order $NK+1$ for modeling the variable delays when $\epsilon=0.001$, $K=27$ and $N' \in \{ 8,32,128 \}$.}
	 \label{fig:hashrates}
\end{figure}

In \figref{fig:hashrates}, we show how mining rate varies over time for both full range data as well as congested network data, i.e., $\alpha(t)$, where we pick $\epsilon=0.01$, $K=27$ and $N' \in \{ 8,32,128 \}$. Notice that, especially in congested network situation, as delays are larger, eventually $T$-block interval rule results in easier puzzles. This in turn means at full honest power, i.e., when everyone receives the most recently mined block, $\alpha>1/600$. Eventually, this will also affect the double spending probability since $\beta$ is a constant fraction of full power $\alpha$ and it increases as well. In other words, during network congestion, as puzzle difficulty drops, adversary benefits from it.

\begin{figure}[t]
	\centering
	\includegraphics[width=0.84\columnwidth]{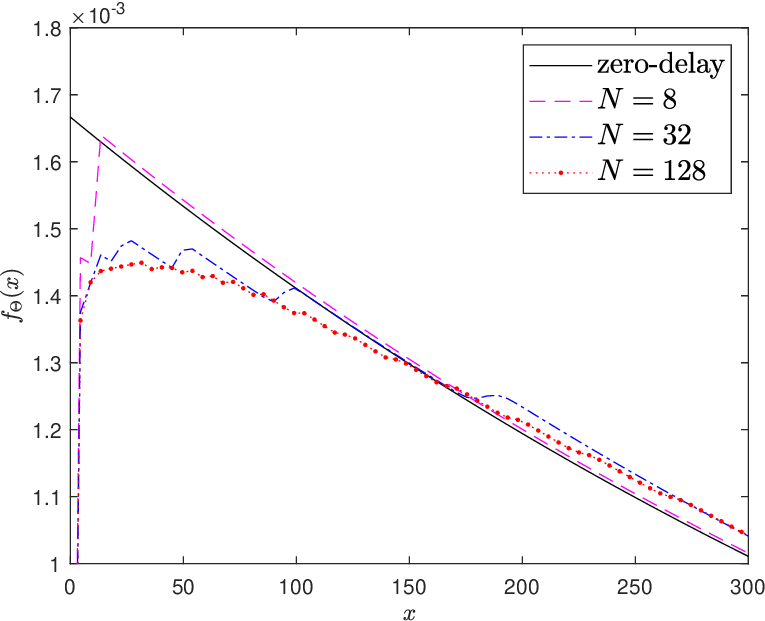}
	\caption{The pdf $f_{\Theta}(x)$ obtained with the zero-delay model and additionally the proposed ME distribution with order $NK+1$ for modeling the variable delays when $\epsilon=0.001$, $K=27$ and $N' \in \{ 8,32,128 \}$.}
	\label{fig:density}
\end{figure}

Next, we compare the pdf of the inter-block mining times, $f_{\Theta}(x)$, of the zero-delay model and the variable-delay model obtained from ME-approximation using the congested network data in \figref{fig:density}. For zero-delay model, we use $\alpha=1/600$. For variable-delay model, we find the $\alpha$ parameter with the iterative method that ensures $\mathbb{E} [\Theta] = 600$ seconds. We then fix the model parameters $\epsilon=0.001$, $K=27$ and $N' \in \{ 8,32,128 \}$. Although the mean inter-block mining times are always 600 seconds, the shape of the corresponding density is different for each case, and as $N'$ increases, we fit the density better to data. For the remaining numerical examples, we will fix $N'$ to 128 for more accurate estimation of the double spend probabilities.

\begin{figure}[t]
	\centering
	\includegraphics[width=0.84\columnwidth]{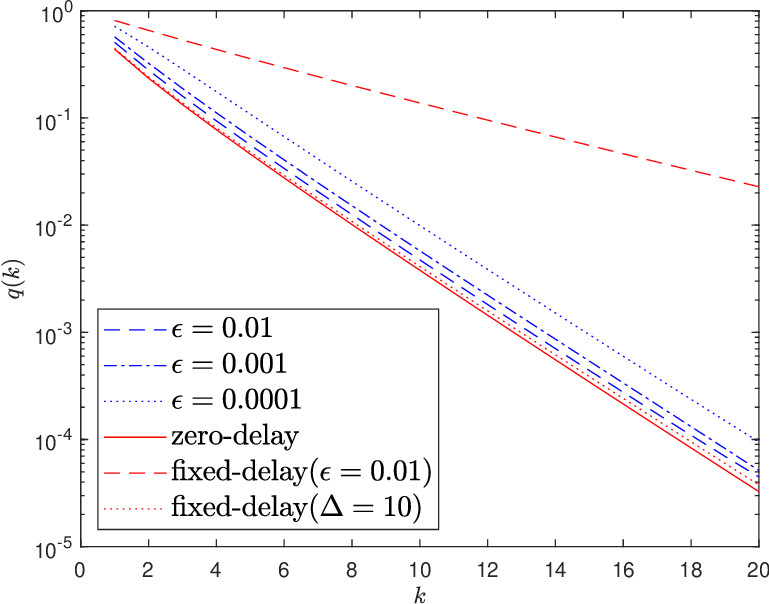}
	\caption{Double spending probability $q$ as a function of $k$ for various delay models.}
	\label{fig:fig3}
\end{figure}

Although we skip plotting the hashrate and density of fixed-delay model, they are fairly easy to obtain by picking $\alpha=\frac{1}{\mathcal{T}-\Delta}$ in exponential distribution and shifting that density by $\Delta$ to the right. However, $\Delta$ is sensitive to the type of data and $\epsilon$ we pick. For example, for full range data, picking  $\epsilon=0.1$ results in $\Delta\approx 8s$ which confirms the assumption that $90$ percentile of the data has delays below $10$ seconds \cite{guo-btc-sec-lat,our-sec-lat-extended,cao2023tradeoff}. On the other hand, picking $\epsilon=0.01$ results in $\Delta\approx 300s$, which mean that some nodes have very high delays and hence can be deceived by the adversary more easily during this time.

In Fig.~\ref{fig:fig3}, we compare the double spend probability $q$ obtained with the zero-delay model as well as the variable-delay model using the proposed ME distribution for three different values of the cutoff parameter $\epsilon \in \{ 0.01, 0.001, 0.0001 \}$, where $\beta=0.2\alpha$ using the full range data. For the fixed-delay model, we plot both the $\Delta=10$ case as in \cite{our-sec-lat-extended}, and additionally pick $\epsilon=0.01$ and find the corresponding $\Delta$ from full range data ($\Delta\approx 300s$). Note that the choice of the cutoff parameter $\epsilon$ has a substantial impact on the double spend probabilities whose accurate estimation is key to the security. Further, the fixed-delay model either ends up almost the same as zero-delay model or gives extremely pessimistic values when only some of the nodes experience high delays.

Note that, the fixed delay model with $\Delta=10$ would correspond to a cutoff parameter of approximately $\epsilon=0.1$ in the delay data we use. The models that used $\Delta=10$ in the literature such as \cite[Section~III.C]{cao2023tradeoff} acknowledge that higher delays are present in real world data but they can be seen as adversarial entities which is a pessimistic assumption. Therefore, it is not fair to compare a scenario of different adversarial powers with our model. Hence, we opted to present two cases of the fixed delay model: one where it is optimistic and ignores the remaining delays ($\epsilon=0.1$) and the other (more pessimistic) where the cutoff is $\epsilon=0.01$. Further, we opted to present Bitcoin results due to the abundance of reliable real world data and considerable interest from the research community and broader public. However, our model could be even more interesting for PoW blockchains that have smaller block interarrival times where the need of a better approximation of double spending probabilities is evident.

\section{Discussion}
In this paper, we introduced a ruin-theoretical analytical model of double spending using a matrix-analytical algorithm by which empirical network delays are used to construct a matrix-exponential distribution to model honest inter-mining durations, when the adversary hashrate is fixed, i.e., the adversary inter-mining periods are exponentially distributed with fixed rate. We proposed a novel method that incorporates the effect of arbitrary network delays on double spending probabilities, which was missing in the literature, even for fixed adversary hashrates. We leave the modeling of varying adversary hashrates for future research.

We used empirical network delays reported in the literature to construct a matrix-exponential distribution to model honest inter-mining durations. While doing so, in our numerical results, we made the assumption that all miners have the same hashrate, which may be a simplistic assumption. This assumption was attributed to the fact that we did not have access to data showing the hash powers associated with inventory messages. It is possible to re-run the model with a more detailed empirical network delay data that reports the hash powers associated with inventory messages.

Our model is heuristic, however, our goal is not to provide pessimistic bounds, but rather lay out a foundational framework, i.e., ruin theory, for more realistic estimation of double spending probabilities in PoW blockchains and more general scenarios. The accuracy depends on the assumption that future network delays follow the same pattern as observed in the past. However, the model can be re-run as new network delay data emerges. Taking into account of a finite collection of miner pools with varying hashrates and detailed peer-to-peer delays within the ruin-theoretical model is an interesting future research direction. We believe that the ruin-theoretical double spending model we introduced in this paper will lay out a foundation for more realistic and interesting blockchain scenarios, more general than that covered by the conventional gambler’s ruin framework. In fact, \eqref{NumEventsRandomTime} can be modified using the iterated law of expectation to consider different pool hashrates and their respective delays to the other pools.

The ME-distribution model allows us to take a unifying view to many other problems studied extensively in the literature as sub-cases, such as zero-delay, fixed-delay, exponentially-fixed-delay, and more specifically variable-delays, the latter being the main scope of this paper. The order $m$ of the characterizing matrices of the ME distribution governs the complexity $\mathcal{O}(m^3)$ of the approach since the algorithm involves the solution of matrix equations of size $m$ which is far larger than the parameter $k$ in the $k$-deep confirmation rule. In the model we propose for variable delays, $m=NK+1$ where $N$ is the discretization parameter and $K$ is the order of the CME distribution. In the numerical examples, we take $K=27$, $N$ up to 130, and the results converged within reasonable accuracy for these parameters, for which we have been able to obtain the double spending probabilities in seconds, with ordinary computing resources.

\bibliographystyle{IEEEtran}
\bibliography{blockchain}

\begin{thebibliography}{10}
\providecommand{\url}[1]{#1}
\csname url@samestyle\endcsname
\providecommand{\newblock}{\relax}
\providecommand{\bibinfo}[2]{#2}
\providecommand{\BIBentrySTDinterwordspacing}{\spaceskip=0pt\relax}
\providecommand{\BIBentryALTinterwordstretchfactor}{4}
\providecommand{\BIBentryALTinterwordspacing}{\spaceskip=\fontdimen2\font plus
\BIBentryALTinterwordstretchfactor\fontdimen3\font minus \fontdimen4\font\relax}
\providecommand{\BIBforeignlanguage}[2]{{%
\expandafter\ifx\csname l@#1\endcsname\relax
\typeout{** WARNING: IEEEtran.bst: No hyphenation pattern has been}%
\typeout{** loaded for the language `#1'. Using the pattern for}%
\typeout{** the default language instead.}%
\else
\language=\csname l@#1\endcsname
\fi
#2}}
\providecommand{\BIBdecl}{\relax}
\BIBdecl

\bibitem{btc-whitepaper}
S.~Nakamoto, ``Bitcoin: A peer-to-peer electronic cash system,'' https://bitcoin.org/bitcoin.pdf, March 2008.

\bibitem{information_propagation}
C.~Decker and R.~Wattenhofer, ``Information propagation in the {Bitcoin} network,'' in \emph{IEEE P2P}, September 2013.

\bibitem{jiang_wu_BC21}
S.~Jiang and J.~Wu, ``Taming propagation delay and fork rate in bitcoin mining network,'' in \emph{2021 IEEE International Conference on Blockchain (Blockchain)}, 2021, pp. 314--320.

\bibitem{bashir2020mastering}
I.~Bashir, \emph{Mastering Blockchain: A Deep Dive into Distributed Ledgers, Consensus Protocols, Smart Contracts, DApps, Cryptocurrencies, Ethereum, and More}.\hskip 1em plus 0.5em minus 0.4em\relax Packt Publishing Ltd, 2020.

\bibitem{xiao2020survey}
Y.~Xiao, N.~Zhang, W.~Lou, and Y.~T. Hou, ``A survey of distributed consensus protocols for blockchain networks,'' \emph{IEEE Communications Surveys \& Tutorials}, vol.~22, no.~2, pp. 1432--1465, 2020.

\bibitem{narayanan2016bitcoin}
A.~Narayanan, J.~Bonneau, E.~Felten, A.~Miller, and S.~Goldfeder, \emph{Bitcoin and Cryptocurrency Technologies: A Comprehensive Introduction}.\hskip 1em plus 0.5em minus 0.4em\relax Princeton University Press, 2016.

\bibitem{rosenfeld2014analysis}
M.~Rosenfeld, ``Analysis of hashrate-based double spending,'' 2014, arXiv:1402.2009.

\bibitem{feller1968introduction}
W.~Feller, \emph{An Introduction to Probability Theory and Its Applications, Volume I}, 3rd~ed.\hskip 1em plus 0.5em minus 0.4em\relax John Wiley \& Sons, 1968.

\bibitem{DoubleSpendRaces}
C.~Grunspan and R.~P\'{e}rez-Marco, ``Double spend races,'' \emph{International Journal of Theoretical and Applied Finance}, vol.~21, no.~8, p. 1850053, 2018.

\bibitem{PINZON201679}
C.~Pinzón and C.~Rocha, ``Double-spend attack models with time advantange for bitcoin,'' \emph{Electronic Notes in Theoretical Computer Science}, vol. 329, pp. 79--103, 2016.

\bibitem{grunspan2020mathematics}
C.~Grunspan and R.~Pérez-Marco, ``The mathematics of bitcoin,'' \emph{European Mathematical Society - Newsletter}, vol. 115, p. 31–37, 2020.

\bibitem{nakamoto-always-wins}
A.~Dembo, S.~Kannan, E.~Tas, D.~Tse, P.~Viswanath, X.~Wang, and O.~Zeitouni, ``Everything is a race and {N}akamoto always wins,'' in \emph{ACM CCS}, November 2020.

\bibitem{garay}
J.~Garay, A.~Kiayias, and N.~Leonardos, ``The bitcoin backbone protocol: Analysis and applications,'' in \emph{EUROCRYPT}, April 2015.

\bibitem{gazi-consist-bounds}
P.~Ga\v{z}i, A.~Kiayias, and A.~Russell, ``Tight consistency bounds for bitcoin,'' in \emph{ACM CCS}, November 2020.

\bibitem{kiffer-method-analyze-consistency}
L.~Kiffer, R.~Rajaraman, and A.~Shelat, ``A better method to analyze blockchain consistency,'' in \emph{ACM CCS}, October 2018.

\bibitem{pass-Analysis-blockchain}
R.~Pass, L.~Seeman, and A.~Shelat, ``Analysis of the blockchain protocol in asynchronous networks,'' in \emph{EUROCRYPT}, May 2017.

\bibitem{guo-close-sec-lat}
J.~Li, D.~Guo, and L.~Ren, ``Close latency-security trade-off for the {N}akamoto consensus,'' in \emph{ACM AFT}, November 2021.

\bibitem{our-sec-lat-extended}
M.~Doger and S.~Ulukus, ``Refined bitcoin security-latency under network delay,'' \emph{IEEE Transactions on Information Theory}, vol.~71, no.~4, pp. 3038--3047, 2025.

\bibitem{cao2023tradeoff}
S.-J. Cao and D.~Guo, ``Trade-off of security, latency, and throughput of the {N}akamoto consensus,'' \emph{IEEE Transactions on Information Theory}, 2025.

\bibitem{gazi-sec-lat}
P.~Ga\v{z}i, L.~Ren, and A.~Russell, ``Practical settlement bounds for proof-of-work blockchains,'' in \emph{ACM CCS}, November 2022.

\bibitem{DSN-Bitcoin-Monitoring}
``{DSN Bitcoin Monitoring},'' \url{https://dsn.tm.kit.edu/bitcoin/}, accessed: 2024-05-05.

\bibitem{pow-under-random-safe}
S.~Sankagiri, S.~Gandlur, and B.~Hajek, ``The longest-chain protocol under random delays.'' available online at arXiv2102.00973.

\bibitem{our-queue-sec-ext-version}
M.~Doger and S.~Ulukus, ``Transaction capacity, security and latency in blockchains,'' available online at arXiv2402.10138.

\bibitem{our-random-delay}
M.~Doger, S.~Ulukus, and N.~Akar, ``Pow security-latency under random delays and the effect of transaction fees,'' in \emph{IEEE ITW}, November 2024.

\bibitem{parallel-pow-bounds}
P.~Keller and R.~B\"{o}hme, ``Parallel proof-of-work with concrete bounds,'' in \emph{ACM AFT}, September 2022.

\bibitem{gap-game}
I.~Tsabary and I.~Eyal, ``The gap game,'' in \emph{ACM CCS}, October 2018.

\bibitem{instability-no-block-reward}
M.~Carlsten, H.~Kalodner, S.~M. Weinberg, and A.~Narayanan, ``On the instability of bitcoin without the block reward,'' in \emph{ACM CCS}, October 2016.

\bibitem{guo-btc-sec-lat}
D.~Guo and L.~Ren, ``Bitcoin's latency–security analysis made simple,'' in \emph{ACM AFT}, September 2023.

\bibitem{Sompolinsky2015SecureHT}
Y.~Sompolinsky and A.~Zohar, ``Secure high-rate transaction processing in bitcoin,'' in \emph{Financial Cryptography}, 2015.

\bibitem{coin-hopping-meshkov}
D.~Meshkov, A.~Chepurnoy, and M.~Jansen, ``Short paper: Revisiting difficulty control for blockchain systems,'' in \emph{Data Privacy Management, Cryptocurrencies and Blockchain Technology}.\hskip 1em plus 0.5em minus 0.4em\relax Springer International Publishing, 2017.

\bibitem{profit_lag}
C.~Grunspan and R.~P{\'e}rez-Marco, ``Profit lag and alternate network mining,'' in \emph{Springer MARBLE}, 2023, pp. 115--132.

\bibitem{asmussen-ruin}
S.~Asmussen, \emph{Ruin Probabilities}.\hskip 1em plus 0.5em minus 0.4em\relax World Scientific, 2000.

\bibitem{Dickson_2016}
D.~C.~M. Dickson, \emph{Insurance Risk and Ruin}, 2nd~ed., ser. International Series on Actuarial Science.\hskip 1em plus 0.5em minus 0.4em\relax Cambridge University Press, 2016.

\bibitem{neuts81}
M.~F. Neuts, \emph{Matrix-geometric Solutions in Stochastic Models: An Algorithmic Approach}.\hskip 1em plus 0.5em minus 0.4em\relax Dover Publications, Inc., 1981.

\bibitem{neuts2013}
------, \emph{Phase-type Probability Distributions}.\hskip 1em plus 0.5em minus 0.4em\relax Boston, MA: Springer US, 2013, pp. 1132--1134.

\bibitem{asmussen_etal_SJS96}
S.~Asmussen, O.~Nerman, and M.~Olsson, ``Fitting phase-type distributions via the em algorithm,'' \emph{Scandinavian Journal of Statistics}, vol.~23, no.~4, pp. 419--441, 1996.

\bibitem{AsmussenBladt97}
S.~Asmussen and M.~Bladt, ``\BIBforeignlanguage{English}{Renewal theory and queueing algorithms for matrix-exponential distributions},'' in \emph{\BIBforeignlanguage{English}{Matrix-analytic methods in stochastic models}}, A.~Alfa and S.~Chakravarthy, Eds.\hskip 1em plus 0.5em minus 0.4em\relax Marcel Dekker, 1996, pp. 313--341.

\bibitem{fackrell_thesis}
M.~W. Fackrell, ``Characterization of matrix-exponential distributions,'' Ph.D. dissertation, The University of Adelaide, 2003.

\bibitem{he_aap07}
Q.-M. He and H.~Zhang, ``\BIBforeignlanguage{English}{On matrix exponential distributions},'' \emph{\BIBforeignlanguage{English}{Advances in Applied Probability}}, vol.~39, no.~1, pp. pp. 271--292, 2007.

\bibitem{Horvath2016}
A.~Horvath, M.~Scarpa, and M.~Telek, \emph{Phase Type and Matrix Exponential Distributions in Stochastic Modeling}.\hskip 1em plus 0.5em minus 0.4em\relax Cham: Springer International Publishing, 2016, pp. 3--25.

\bibitem{Bean2010}
N.~Bean and B.~F. Nielsen, ``Quasi-birth-and-death processes with rational arrival process components,'' \emph{Stochastic Models}, vol.~26, pp. 309--334, 07 2010.

\bibitem{BUCH09a}
P.~Buchholz and M.~Telek, ``Stochastic {P}etri nets with matrix exponentially distributed firing times,'' \emph{Performance Evaluation}, vol.~67, pp. 1373--1385, 2010.

\bibitem{BUCH10b}
------, ``Rational processes related to communicating {M}arkov processes,'' \emph{Journal of Applied Probability}, vol.~49, pp. 40--59, 2012.

\bibitem{asmussen.erlangian.2002}
S.~Asmussen, F.~Avram, and M.~Usabel, ``Erlangian approximations for finite-horizon ruin probabilities,'' \emph{Astin Bulletin}, vol.~32, no.~2, pp. 267--282, 2002.

\bibitem{ramaswami.2008}
V.~Ramaswami, D.~G. Woolford, and D.~A. Stanford, ``The {E}rlangization method for {M}arkovian fluid flows,'' \emph{Annals of Operations Research}, vol. 160, no.~1, pp. 215--225, 2008.

\bibitem{cme}
I.~Horv{\'a}th, O.~S{\'a}f{\'a}r, M.~Telek, and B.~Z{\'a}mb{\'o}, ``Concentrated matrix exponential distributions,'' in \emph{Computer Performance Engineering}, D.~Fiems, M.~Paolieri, and A.~N. Platis, Eds.\hskip 1em plus 0.5em minus 0.4em\relax Cham: Springer International Publishing, 2016, pp. 18--31.

\bibitem{MEfication}
N.~Akar, O.~Gursoy, G.~Horvath, and M.~Telek, ``Transient and first passage time distributions of first- and second-order multi-regime {Markov} fluid queues via {ME-fication},'' \emph{Methodology and Computing in Applied Probability}, vol.~23, pp. 1257--1283, 2021.

\bibitem{yates2014probability}
R.~Yates and D.~Goodman, \emph{Probability and Stochastic Processes: A Friendly Introduction for Electrical and Computer Engineers}.\hskip 1em plus 0.5em minus 0.4em\relax Wiley, 2014.

\bibitem{alfa2015applied}
A.~Alfa, \emph{Applied Discrete-Time Queues}.\hskip 1em plus 0.5em minus 0.4em\relax Springer New York, 2015.

\bibitem{Lindley_1952}
D.~V. Lindley, ``The theory of queues with a single server,'' \emph{Mathematical Proceedings of the Cambridge Philosophical Society}, vol.~48, no.~2, p. 277–289, 1952.

\bibitem{goffard_insurance_selfish}
A.~Hansjoerg and P.-O. Goffard, ``On the profitability of selfish blockchain mining under consideration of ruin,'' \emph{Operations Research}, vol.~70, no.~1, pp. 179--200, 2022.

\bibitem{Goffard_Fraud_risk}
P.-O. Goffard, ``Fraud risk assessment within blockchain transactions,'' \emph{Advances in Applied Probability}, vol.~51, no.~2, p. 443–467, 2019.

\bibitem{sompolinsky2016bitcoin}
\BIBentryALTinterwordspacing
Y.~Sompolinsky and A.~Zohar, ``Bitcoin's security model revisited,'' 2016. [Online]. Available: \url{https://arxiv.org/abs/1605.09193}
\BIBentrySTDinterwordspacing

\end{thebibliography}
\end{document}